\documentclass[twocolumn,showpacs,amsmath,amssymb]{revtex4}

\usepackage{graphicx}
\usepackage{dcolumn}
\usepackage{bm}
\usepackage{psfig}  
\usepackage{epsfig}

\begin{document}
\title{Nuclear Shadowing in Neutrino-Nucleus Deeply Inelastic Scattering}

\author{Jianwei Qiu}%

\author{Ivan Vitev}

\affiliation{Department of Physics and Astronomy, 
Iowa State University, Ames, IA 50011, USA }

\begin{abstract}
In the framework of the collinear factorized 
pQCD approach 
we calculate the small-$x_B$ process-dependent nuclear modification to
the structure functions measured in neutrino-nucleus deeply inelastic 
scattering. We include both heavy quark mass corrections $(M^2/Q^2)$ and 
resummed nuclear-enhanced dynamical power corrections 
in the quantity $(\xi^2/Q^2)(A^{1/3}-1)$ with 
$\xi^2$ evaluated to leading order in $\alpha_s$.  
Our formalism predicts a measurable  difference in the  shadowing 
pattern of the structure functions $F_2^A(x_B,Q^2)$ and
$F_3^A(x_B,Q^2)$ and a significant low- and moderate-$Q^2$ modification 
of the QCD sum rules.
We also comment on the relevance of our results to the NuTeV 
extraction of $\sin^2\theta_W$.
\end{abstract}
                                               

\maketitle

\section{Introduction}

Recent surprising results on $\sin^2 \theta_W$,  
reported by the NuTeV collaboration and based on a comparison of 
charged and neutral current neutrino interactions with an iron rich 
target~\cite{Zeller:2001hh}, 
renewed our quest for understanding the nuclear dependence in 
neutrino-nucleus deeply inelastic scattering (DIS).  
A possibility that process-dependent nuclear shadowing might
affect the NuTeV  extraction of the Weinberg angle $\theta_W$ 
was raised by Miller and Thomas~\cite{Miller:2002xh}.
Although such scenario was considered unlikely 
by the  collaboration~\cite{Zeller:2002et}, a systematic study and 
a clear understanding of the process-dependent nuclear effects 
in neutrino-nucleus scattering will strengthen the importance of the 
NuTeV result.

Like all nuclear dependences in the physical cross sections 
\cite{Qiu:2001hj}, 
the small-$x_B$ shadowing in lepton-nucleus DIS has both 
process-dependent and process-independent contributions.  
While its universal part can be factorized in the 
leading twist nuclear parton distribution functions (nPDFs), 
the DIS-specific  modifications arise from the higher twist 
(or power) corrections to 
the structure functions~\cite{Miller:2002xh,Qiu:2002mh,Qiu:2003vd}. 
In this letter, 
we present a calculation of the process-dependent 
shadowing  
in neutrino-nucleus 
deeply inelastic scattering by resumming heavy quark mass corrections, 
$M^2/(2p\cdot q)=x_B M^2/Q^2$, and 
nuclear size enhanced dynamical power corrections, 
$(\xi^2/Q^2)(A^{1/3}-1)$ with $\xi^2 \propto 
\langle F^{+\perp}F^{+}_{\ \perp}\rangle$, the gluon density in a large
nucleus. The numerical value for the characteristic scale of 
higher twist $\xi^2$  \cite{Qiu:2003vd}, extracted from DIS data
on $\mu$-$A$ interactions \cite{Arneodo:1995cs}, 
is much less than  $Q^2$ in the region  which 
is perturbatively accessible. 
Therefore, we only evaluate $\xi^2$ to the leading order in $\alpha_s$, 
while resumming the power corrections to all orders in 
$(\xi^2/Q^2)(A^{1/3}-1)$.  Using $\xi^2=0.09-0.12$~GeV$^2$ 
\cite{Qiu:2003vd}, our results provide a good description of the 
deviation between  the Gross-Llewellyn Smith 
QCD sum rule \cite{Gross:1969jf} adjusted for ${\cal O}(\alpha_s)$ scaling 
violations  \cite{Brock:1993sz} and the existing 
data \cite{Kim:1998ki}. At small Bjorken $x_B$, the high 
twist components to the calculated structure functions 
$F_2^{A}(x_B,Q^2)$ and $F_3^{A}(x_B,Q^2)$ 
in neutrino-iron DIS  qualitatively describe the low-$x_B$ 
and low-$Q^2$ suppression trend in the preliminary data, 
recently reported by the 
NuTeV collaboration at DIS 2003 \cite{Naples:2003ne}.

In the next Section we briefly review the DIS kinematics and 
coherence at small Bjorken $x_B$.  
In Section III we demonstrate  that at the tree level
mass corrections and  dynamical power corrections 
``commute'' and their resummation can be carried out in a closed form.  
We derive analytic expressions for the process-dependent 
nuclear modification to the transverse and longitudinal structure 
functions in neutrino-nucleus DIS. 
In Section~IV we predict the difference in the shadowing pattern
of $F_2^A(x_B,Q^2)$  and  $F_3^A(x_B,Q^2)$, and   
give quantitative results for the $x_B$-, $A$- and $Q^2$-dependence 
of the nuclear modification to the charged current $\nu(\bar{\nu})$-$A$ 
DIS structure functions. 
We find sizable small- and moderate-$Q^2$ corrections to 
the Gross-Llewellyn  Smith  
QCD sum rule.
In Section~V we comment on the relevance of our results to the NuTeV 
extraction of $\sin^2\theta_W$.  
Finally, we give our conclusions in Section~VI.

\section{DIS Kinematics and Coherence at Small $x_B$}

\begin{figure}[t!]
\begin{center} 
\hspace*{-.1in}\psfig{file=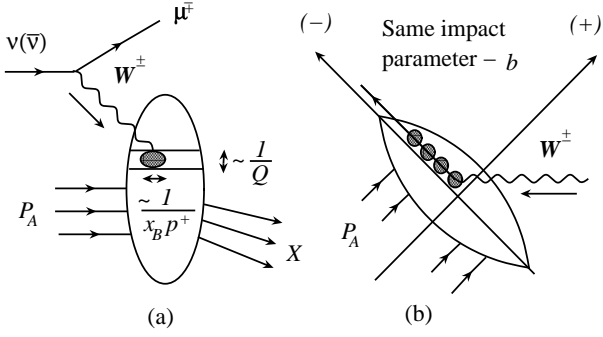,height=2.in,width=3.7in,angle=0}
\vspace*{-.3in}
\caption{(a) Illustration of the characteristic scales, 
$A_\perp = 1 / Q^2$ and $\Delta z^{(-)} = 1/(x_B p^+)$ in the 
boosted frame, probed by the virtual meson in DIS.  
(b)  Multiple final state interactions of the struck quark with 
the partons from the nucleus at a fixed impact parameter $b$.}
\label{DIS-kin}
\end{center} 
\vspace*{-4mm}
\end{figure}

The charged current 
DIS cross section of a neutrino (or antineutrino) beam ($k$) 
off a nuclear target ($P_A$), 
as illustrated in Fig.~\ref{DIS-kin}(a), 
probes three independent structure functions, 
$F_i^A(x_B,Q^2)$ with $i=1,2,3$ \cite{Brock:1993sz}
\begin{eqnarray}
\frac{d\sigma^{\nu(\bar{\nu})A}}{dx_B dy}
&=& \frac{\pi\alpha^2_{\rm em} m_N E}{2\sin^4(\theta_W)(Q^2+M_W^2)^2}
\nonumber\\
&\times &
\bigg[ \frac{y^2}{2} 2x_B F_1^{\nu(\bar{\nu})A}
      +\left(1-y-y\frac{m_N x_B}{2E}\right) F_2^{\nu(\bar{\nu})A}
\nonumber\\
&\ & + (-) 
         \left(y-\frac{y^2}{2}\right) 
         x_B F_3^{\nu(\bar{\nu})A} \bigg]\, ,
\label{diffdis}
\end{eqnarray}
where the Bjorken variable $x_B= Q^2/(2p\cdot q)$ with $p = P_A/A$, 
the exchanged $W$-boson momentum $q$ and its virtuality $Q^2=-q^2$, 
and $y=p\cdot q/(p\cdot k)$.
In Eq.~(\ref{diffdis}), the ``$(-)$'' represents the sign for 
an antineutrino beam, $m_N=M_A/A$ with nuclear mass $M_A$,
$M_W$ is the $W$-boson mass, and $E$ is the beam energy. 
The often-referred longitudinal structure function,
$F_L^{\nu(\bar{\nu})A} = F_2^{\nu(\bar{\nu})A}/(2x_B) - 
F_1^{\nu(\bar{\nu})A}$ if $4x_B^2m_N^2\ll Q^2$ \cite{Brock:1993sz}. 
Here, we are predominantly interest in the small-$x_B$ region
and neglect the target mass (rescaling) 
corrections~\cite{Georgi:1976ve}.

The DIS cross section with an exchange of a $W$- or $Z$-boson 
of virtuality $Q^2$
and energy $\nu = Q^2/(2m_Nx_B)$ has an effective resolution 
in transverse area $A_\perp = 1 / Q^2$, which is much less than
the nucleon size, and an uncertainty in longitudinal direction 
$\Delta z^{(-)} = 1/(x_Bp^{+})$ with boosted nucleon momentum $p^{+}$. 
If $\Delta z^{(-)} = 1/(x_Bp^{+}) \geq 2r_0(m_N/p^{+})$ or 
$x_B \leq x_N = 1/(2m_N r_0) \sim 0.1$, the neutrino will 
{\it coherently} interact with more than one nucleon inside the nucleus, 
and probe the nuclear dependence at a perturbative scale $Q^2$ 
\cite{Qiu:2002mh,Qiu:2003vd,Mueller:1999wm}.

\section{Calculating mass and dynamical power corrections}

\begin{figure}[t!]
\begin{center} 
\psfig{file=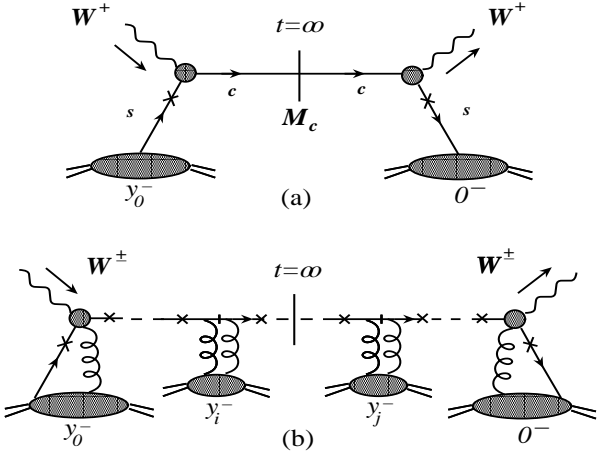,height=2.6in,width=3.5in,angle=0}
\vspace*{-.3in}
\caption{(a) Tree level direct coupling of the exchange vector 
meson $W^\pm$ to the struck quark in charge current 
neutrino-nucleus DIS  with massive charm final state.  
(b) Tree-level contribution to the nuclear-enhanced dynamical  
power corrections with heavy quark effects.}
\label{fig1:F-rule}
\end{center} 
\vspace*{-4mm}
\end{figure}

Electroweak charged and neutral current  processes necessitate a 
discussion of final state charm mass effects in neutrino-nucleus DIS  
even if the leading twist charm quark parton distribution is neglected, 
$\phi_c(x,Q^2)=\phi_{\bar{c}}(x,Q^2)=0$ \cite{Barnett:1976}.  
It is, therefore, critical to develop a systematic approach to 
the interplay of a heavy quark final state and the resummed nuclear 
enhanced power corrections discussed in~\cite{Qiu:2003vd}. We 
define the boost invariant mass fraction 
\begin{equation}
x_M = \frac{M^2}{ 2 p\cdot q} = x_B \, \frac{M^2}{Q^2}
\end{equation}   
and choose a frame such that $p^\mu=p^+\bar{n}^\mu$ and
$q^\mu = -x_Bp^+ \bar{n}^\mu + Q^2/(2x_Bp^+)n^\mu$, where
$\bar{n}^\mu = [1,0,0_\perp]$ and $n^\mu = [0,1,0_\perp]$  
specify the ``+'' and ``$-$'' lightcone directions, respectively.
With a non-vanishing quark mass $M$, the Feynman rule for the  
final state cut line of quark momentum $x_ip+q$ 
in Fig.~\ref{fig1:F-rule}(a) is
\begin{equation} 
{\rm Cut} = 2\pi \left( \frac{x_B}{Q^2} \right) 
 \left(  \gamma\cdot\tilde{p} + M  \right) \delta(x_i-x_B-x_M) \, , 
\label{cut}
\end{equation}
where 
\begin{equation}
\tilde{p}^\mu= x_M p^+\, \bar{n}^\mu 
             + \frac{Q^2}{2 x_B p^+}\, n^\mu \; ,
\label{ptilde}
\end{equation}
with $\tilde{p}^2 = M^2$.
For $M\rightarrow 0$ we recover the known massless case, in which 
the scattered quark is moving along the ``$-$'' lightcone direction.  
A direct consequence of this Feynman rule is a tree-level
coupling for longitudinally polarized vector mesons $\propto M^2/Q^2$.
Contracting $\epsilon_L^{\mu\nu}$~\cite{Guo:2001tz} with the charged current 
hadronic tensor $W_{\mu\nu}$~\cite{Brock:1993sz} yields:
\begin{eqnarray} 
\frac{1}{A} 
F_L^{\nu A} (x_B,Q^2) &=&  
\sum_{D,U}   |V_{DU}|^2 \,
 \frac{{M_U^2}}{Q^2} \,
\phi_D^A \left(x_B + x_{M_U},Q^2 \right) \nonumber  \\
&\ & {\hskip -0.6in}
+  \,  \sum_{\bar{U},\bar{D}} |V_{\bar{U} \bar{D}}|^2 \,
 \frac{{M_{\bar{D}}^2}}{Q^2} \,
\phi_{\bar{U}}^A \left(x_B + x_{M_{\bar{D}}},Q^2 \right)   \, , \quad
\label{FLmassWp}  \\ 
\frac{1}{A} 
F_L^{\bar{\nu} A} (x_B,Q^2) & = &
   \sum_{U,D} |V_{UD}|^2 \, \frac{{M_D^2}}{Q^2} \,
\phi_U^A \left(x_B + x_{M_D},Q^2 \right) \nonumber  \\
&\ & {\hskip -0.6in}
+ \,   \sum_{\bar{D},\bar{U}} |V_{\bar{D} \bar{U}}|^2 \, 
  \frac{{M_{\bar{U}}^2}}{Q^2} \,
\phi_{\bar{D}}^A \left(x_B + x_{M_{\bar{U}}},Q^2 \right)   \, ,
\label{FLmassWn}  
\end{eqnarray}
where the CKM matrix elements $V_{ij}$ parametrize the 
electroweak and mass eigenstate mixing with 
up-type quark $U=(u,c,t)$ and down-type quark $D=(d,s,b)$ 
\cite{Brock:1993sz}, and  
$\phi_i^A$ represent the flavor-$i$ universal twist-2 parton 
distribution functions (PDFs) of a nucleon ($A=1$) or 
a nucleus~\cite{Qiu:2002mh}. 
Eqs.~(\ref{FLmassWp}) and  (\ref{FLmassWn}) give 
a novel leading order ($\alpha_s^0$) power suppressed ($M^2/Q^2$) 
quark mass contribution to 
the ratio of longitudinal and transverse structure functions 
$R(x_B,Q^2) = F_L^A(x_B,Q^2)/F_1^A(x_B,Q^2)$ for both nucleons 
and nuclei.  
%
%

We calculate the nuclear enhanced dynamical power corrections in
the lightcone $A^+=0$ gauge.  In this gauge, other than 
the initial-state contact-term contributions, 
all leading order nuclear enhanced power corrections are from 
final-state multiple gluon interactions of the scattered quark 
in a large nucleus shown in Fig.~\ref{fig1:F-rule}(b) \cite{Qiu:2003vd}.
To resum all order nuclear enhanced power corrections with a non-vanishing 
(anti)quark mass, we examine its propagator structure~\cite{Qiu:1988dn}. 
For a quark momentum $x_ip + q$ 
\begin{eqnarray} 
{\rm Propagator} &  =  &
 \pm i \left(\frac{x_B }{Q^2}\right) 
\gamma\cdot p \nonumber  \\ 
&\ & 
 \pm i \left(\frac{x_B}{Q^2} \right)
\frac{\gamma\cdot\tilde{p} + M }
{x_i-(x_B+x_M) \pm i\epsilon} \;,  \qquad
\label{specprop} 
\end{eqnarray} 
where $\pm i, \pm i\epsilon$ correspond to propagators to the 
left or right of the $t=\infty$ cut. In the Fourier space conjugate 
to $x_i p^+$ the first term, free of $x_i$ pole,  is 
$\propto \delta(y_i^-)$.  The operators in the hadronic matrix element 
that this contact term ($ \longrightarrow  \hspace*{-0.37cm}
 \mid \hspace*{0.3cm} $)  separates can  be evaluated in
the same nucleon state~\cite{Qiu:2003vd}.  
In contrast, the Fourier transform of the 
second term is $\propto \theta(y_i^-)$.  Therefore, this 
pole term  ($ \longrightarrow  
\hspace*{-0.6cm} \times \hspace*{0.25cm}$) is the  source of 
the $A^{1/3}$ nuclear size enhancement to the power corrections. 
The operators that it connects in the multi-field multi-local hadronic 
matrix element can be long-distance separated and thus approximately 
evaluated in different nucleon 
states~\cite{Qiu:2003vd}. Alternative operator 
decompositions as well as other terms that arise from a 
formal operator product expansion (OPE) \cite{Shuryak:1981pi}  
are suppressed by powers of the nuclear size.

The case of massive final state quarks could be much 
more involved than the $M\rightarrow 0$ limit. 
The complexity of the calculation  stems from
the potentially dangerous exponential growth of the number of terms 
coming  from products of propagators, see Fig.~\ref{fig1:F-rule}(b). 
In our calculation we first observe that 
$\gamma\cdot\tilde{p}+M$ in the cut line, Eq.~(\ref{cut}), and 
the numerator of the pole-term, Eq.~(\ref{specprop}), arise from 
an {\it on-shell} momentum $\tilde{p}$. 
Since the exchange gluons at the vertices connected 
by quark propagators are transversely polarized
in the $A^+=0$ physical gauge, 
$A_\mu(y^-_i)\gamma^\mu \approx  A_\perp(y^-_i) \gamma^\perp$,  
$$ \cdots \gamma \cdot p \, \gamma^\perp \, \gamma \cdot p  \cdots 
\propto - p^2 \,  \gamma^\perp = 0 \, , $$
\vspace*{-.7cm}
$$\cdots (\gamma \cdot \tilde{p} + M) \,  \gamma^\perp \,   
(\gamma \cdot \tilde{p} + M) \cdots  \propto 
 - (\tilde{p}^2 - M^2) \,  \gamma^\perp
=   0 \;. $$
For the diagrams in Fig.~\ref{fig1:F-rule}(b)
only one {\em alternating} sequence  of short and long 
distance parts of the propagators Eq.~(\ref{specprop}), 
initiated by the $t=\infty$ cut, survives. 
Therefore, there must be an {\it even} number of gluon interactions 
between the cut line and any surviving pole term of a propagator
in Fig.~\ref{fig1:F-rule}(b).
We also note that
\begin{equation} 
\cdots \gamma\cdot p \, ( \gamma \cdot \tilde{p} \pm M  ) \, 
\gamma\cdot p \cdots    \propto  
\gamma^- p^+\, \left( \frac{Q^2}{2 x_B p^+}\gamma^+ \right) \,   
\gamma^- p^+
\nonumber
\end{equation}
leaves no mass dependence in the spinor trace of the diagrams. 
The 
basic unit for two-gluon exchange with a net momentum fraction flow 
$x_i - x_{i-1}$ and two-quark-propagators 
(one contact plus one pole)~\cite{Qiu:2003vd}, 
see Fig.~\ref{fig1:F-rule}(b), now reads:
\begin{eqnarray}
\label{ScalVert}
&& \!\!\!\!\!\!
{\rm Unit} = x_B \left( \frac{ 4 \pi^2 \alpha_s}{3Q^2}  \right)    
\int  \frac{d \lambda_i}{2 \pi}  \,  \frac{e^{i(x_i-x_{i-1})\lambda_i} }
{x_i - x_{i-1} - i\epsilon} \nonumber \\[1ex]  
&& \qquad \qquad 
\times \left\{
\begin{array}{ll}
  \frac {\gamma^-  \gamma^+}{2} 
\frac{ -i \hat{F}^2(\lambda_i) }{x_{i-1}-(x_B+x_M) +i \epsilon }\;, 
\quad  & {\rm left}  \qquad  \\[1ex]
 \frac { \gamma^+  \gamma^-}{2} 
\frac{ -i \hat{F}^2(\lambda_i) }{x_{i}-(x_B+x_M) -i \epsilon } \;, 
\quad   & {\rm right} \qquad
\end{array}
\right. \;  . 
\end{eqnarray}
In Eq.~(\ref{ScalVert}) the boost invariant 
$\lambda_i = p^+ y_i^-$, the two cases 
correspond to a vertex to the left or right of the final state cut  and  
$\hat{F}^2(\lambda_i)$ is given by the intra-nucleon two-gluon field 
strength correlator defined in~\cite{Qiu:2003vd}:
\begin{equation}
\hat{F}^2 (\lambda_i) \equiv  
\int \frac{d  \tilde{\lambda}_i}{2 \pi} \;  \frac{1}{(p^+)^2}
 F^{+ \alpha}(\lambda_i)  F_{\alpha}^{\; +}(\tilde{\lambda_i}) 
\, \theta(\lambda_i  - \tilde{\lambda_i} ) \;.
\label{FFlambda}
\end{equation}
We conclude that the dynamical nuclear-enhanced 
all twist contributions from the leading order in $\alpha_s$ 
Feynman diagrams with a
massive quark final state are identical to the massless case
up to the substitution $x_B \rightarrow x_B + x_M $ (rescaling) 
in the  $\delta$-function in the cut, Eq.~(\ref{cut}), and 
the propagator poles, Eq.~(\ref{ScalVert}).
%
%
Effectively, we have shown that the mass and nuclear enhanced power 
corrections ``commute'' and $x_i=x_B+x_M$ for all ``$i$''. 
This allows us to take all possible final state interaction diagrams 
and all possible cuts \cite{Qiu:2003pm}  to explicitly carry out 
the resummation of 
coherent high-twist contributions to neutrino-nucleus DIS 
structure functions,
\begin{widetext}

\begin{eqnarray} 
\frac{1}{A} F_{1,3}^{\nu A} (x_B,Q^2) & \approx & \{2\} 
    \left( \sum_{D,U} |V_{DU}|^2 
\phi_D^A \left(x_B + x_{\rm HT} + x_{M_U},Q^2\right)   
   \pm   \sum_{\bar{U},\bar{D}} |V_{\bar{U} \bar{D}}|^2 
\phi_{\bar{U}}^A \left(x_B + x_{\rm HT} + x_{M_{\bar{D}}},Q^2 \right) 
 \right) \, ,  
\label{F13resWp}  \\ 
\frac{1}{A} F_{1,3}^{\bar{\nu} A} (x_B,Q^2) &\approx &  \{2\} 
\left(  \sum_{U,D} |V_{UD}|^2 
\phi_U^A \left(x_B + x_{\rm HT} + x_{M_D},Q^2 \right) 
 \pm  \sum_{\bar{D},\bar{U}} |V_{\bar{D} \bar{U}}|^2 
\phi_{\bar{D}}^A \left(x_B + x_{\rm HT} + x_{M_{\bar{U}}},Q^2 \right) 
\right)   \, .
\label{F13resWn}  
\end{eqnarray}

\end{widetext}

In Eqs.~(\ref{F13resWp}) and (\ref{F13resWn}) the ``$\pm$'' signs 
refer to $F_1$ (parity conserving) and $F_3$ (parity violating) 
transverse structure functions, respectively. 
The factor ``$\{2\}$'' gives the standard normalization 
for $F_3$ only~\cite{Brock:1993sz} and 
the isospin average in the PDFs over the 
protons and neutrons in the nucleus is implicit. 
In Eqs.~(\ref{F13resWp}) and (\ref{F13resWn}) 
$x_{\rm HT}$ 
is the momentum fraction shift (rescaling) 
induced by nuclear enhanced dynamical power corrections and 
derived in Ref.~\cite{Qiu:2003vd}: 
\begin{eqnarray}
x_{\rm HT} &=&
 x_B \, \frac{\xi^2}{Q^2} ({A}^{1/3}-1) \,f(x_B) \; ,    
\label{htx}
\end{eqnarray}
where $\xi^2$ represents the effective scale for the dynamical 
power corrections.  To the leading order in $\alpha_s$
it is given by
\begin{eqnarray}
\xi^2  & = & 
\frac{3 \pi \alpha_s(Q^2)}{8\, r_0^2}
  \langle \, p \, | \hat{F}^2 | \, p \, \rangle     \; ,
\label{xi2}
\end{eqnarray}
where $\langle \, p \, | \hat{F}^2 | \, p \, \rangle$ 
depends on the small-$x$ limit of the 
gluon distribution in the nucleon/nucleus \cite{Qiu:2003vd}.  
While $x_N \approx 0.1$ is the limiting value 
for the onset of coherence,  
at $x_A = 1 / (2 m_N r_0 A^{1/3}) < x_N$ 
the exchange vector meson already 
probes the full nuclear size, see Fig.~\ref{DIS-kin}. 
To first approximation, the function 
\begin{eqnarray}
\label{uncertain}
f(x_B) & = & 
\left\{ \begin{array}{ll}  0\, ,  & x_B > x_N  \\[2ex]
 \frac{ x_B^{-1} - x_N^{-1}}{ x_A^{-1} - x_N^{-1} } \, , &  
 x_A \leq x_B  \leq x_N \\[2ex] 
 1\, , & x_B < x_A   
\end{array}    \right. \;  
\end{eqnarray}
in Eq.~(\ref{htx}) represents the interpolation between the two regimes 
based on the uncertainty principle~\cite{Qiu:wh}.
The nuclear enhancement factor $(A^{1/3}-1)$ in Eq.~(\ref{htx}) 
comes from the integration $\int d\lambda_i$ in Eq.~(\ref{ScalVert}), 
the lower limit of which was chosen such that the effect vanishes for the 
proton ($A=1$) case.

Including the dynamical power corrections, 
the longitudinal structure functions 
in Eqs.~(\ref{FLmassWp}) and (\ref{FLmassWn}) become:
\begin{widetext}

\begin{eqnarray} 
\frac{1}{A} F_L^{\nu A} (x_B,Q^2) 
& \approx & F_L^{({\rm LT})} (x_B,Q^2)
 +         \sum_{D,U}   |V_{DU}|^2
\left[  \frac{{M_U^2}}{Q^2} +  \frac{\xi^2}{Q^2} 
\left( 2 - \frac{{M_U^2}}{Q^2+{M_U^2} } \right)^2  \right] 
\phi_D^A \left(x_B + x_{\rm HT} + x_{M_U},Q^2 \right) \nonumber  \\
&\ & 
+ \,  \sum_{\bar{U},\bar{D}} |V_{\bar{U} \bar{D}}|^2 
\left[  \frac{{M_{\bar{D}}^2}}{Q^2} +  \frac{\xi^2}{Q^2}
\left( 2 - \frac{{M_{\bar{D}}^2}}{Q^2+{M_{\bar{D}}^2} } \right)^2  \right] 
\phi_{\bar{U}}^A \left(x_B + x_{\rm HT} + x_{M_{\bar{D}}},Q^2 \right)   
\, , 
\label{FLresWp}  \\ 
\frac{1}{A} F_L^{\bar{\nu} A} (x_B,Q^2)  
& \approx & F_L^{({\rm LT})} (x,Q^2)
+   \sum_{U,D} |V_{UD}|^2 \left[  \frac{{M_D^2}}{Q^2} +  \frac{\xi^2}{Q^2} 
\left( 2 - \frac{{M_D^2}}{Q^2+{M_D^2} } \right)^2  \right] 
\phi_U^A \left(x_B + x_{\rm HT} + x_{M_D},Q^2 \right) \nonumber  \\
&\ &  
+ \,   \sum_{\bar{D},\bar{U}} |V_{\bar{D} \bar{U}}|^2 
\left[  \frac{{M_{\bar{U}}^2}}{Q^2} +  \frac{\xi^2}{Q^2}
\left( 2 - \frac{{M_{\bar{U}}^2}}{Q^2+{M_{\bar{U}}^2} } \right)^2  \right] 
\phi_{\bar{D}}^A \left(x_B + x_{\rm HT} + x_{M_{\bar{U}}},Q^2 \right)   \, .
\label{FLresWn}  
\end{eqnarray}

\end{widetext}
In Eqs.~(\ref{FLresWp}) and (\ref{FLresWn}) we include the 
${\cal O}(\alpha_s)$ leading twist longitudinal structure functions
$F_L^{({\rm LT})}(x,Q^2)$~\cite{Brock:1993sz} since they are of 
the same order as the leading $\xi^2$ power. 
For numerical evaluation in the next Section 
we consider two quark generations, $U=(u,c)$ and $D=(d,s)$, and
use $|V_{ud}|^2 = |V_{cs}|^2 = \cos^2 \theta_c = 0.95$, 
$|V_{us}|^2 = |V_{cd}|^2 = \sin^2 \theta_c = 0.05$ with 
Cabibbo angle $\theta_c$~\cite{Cabibo}.
The $u,d$ and $s$ quarks  are treated as massless and 
the charm quark mass is set to $M_c=1.35$~GeV \cite{Zeller:2001hh}.

\section{High twists, shadowing  and QCD sum rule }

We first quantify analytically the differences in the ``shadowing''
pattern induced by valance and sea quarks, neglecting the charm mass 
effects that are shown to be small below. For isoscalar-corrected  
($Z=N=A/2$) target nuclei we average over neutrino- and 
antineutrino-initiated  charged  current  interactions, 
$F_i^A(x_B,Q^2)=\big( F_i^{\nu A}(x_B,Q^2) +
F_i^{\bar{\nu} A}(x_B,Q^2)\big)/2$. In the leading-order
and leading twist parton model $F_3^A(x_B,Q^2)$ measures the valance
quark number density with $\phi_{val.}(x) \propto x^{-\alpha_{val.}}$ 
at small $x$. $F_2^A(x_B,Q^2)$, a singlet distribution, 
is proportional to the momentum density of all interacting 
quark constituents and for $x_B \ll 0.1$ is dominated by 
the sea contribution, $\phi_{sea}(x) \propto x^{-\alpha_{sea}}$.
Therefore, the $x_B$-dependent shift
from dynamical nuclear enhanced power corrections, 
$x_{\rm HT}$ in Eq.~(\ref{htx}), generates different
modification to $F_2^A(x_B,Q^2)$ and $F_3^A(x_B,Q^2)$.
Let $R^{A/A^\prime}_{sea/val.}(x_B,Q^2)$ be the shadowing ratio determined
from nuclei $A$ and $A^\prime$ (for example  $^{56}Fe$ to $^2D$)  in
$F_2$ and $F_3$.  If the scale of high twist corrections 
$\xi^2 \ll Q^2$~\cite{Qiu:2003vd} and $x_B \leq \min  
(x_A,x_{A^\prime})$, we expand the PDFs to first order in 
$x_{\rm HT}$ 
 to obtain:
\begin{eqnarray}
&& \!\!\!\!\!\!\!\!\!\!\!\!
R^{A/A^\prime}_{sea/val.}(x_B,Q^2) = \left.
\frac{F_2^{A}(x_B,Q^2)}{F_2^{A^\prime}(x_B,Q^2)} \right/
\frac{F_3^{A}(x_B,Q^2)}{F_3^{A^\prime}(x_B,Q^2)} \nonumber \\[1ex]
 &&= 1-(\alpha_{sea} - \alpha_{val.}) (A^{1/3}-A^{\prime \, 1/3})
 \xi^2/Q^2 + \cdots  \;\; 
\qquad
\label{analyt}
\end{eqnarray} 
Since  $ \alpha_{val.} \approx 0.5 $ and $\alpha_{sea} \approx 1$ vary 
slowly with $Q^2$, Eq.~(\ref{analyt}) predicts a measurable 
difference in the nuclear shadowing  for the structure function
$F_2$ ($F_1$) in comparison to $F_3$.

\begin{figure}[t!]
\begin{center} 
\vspace*{.3in}
\psfig{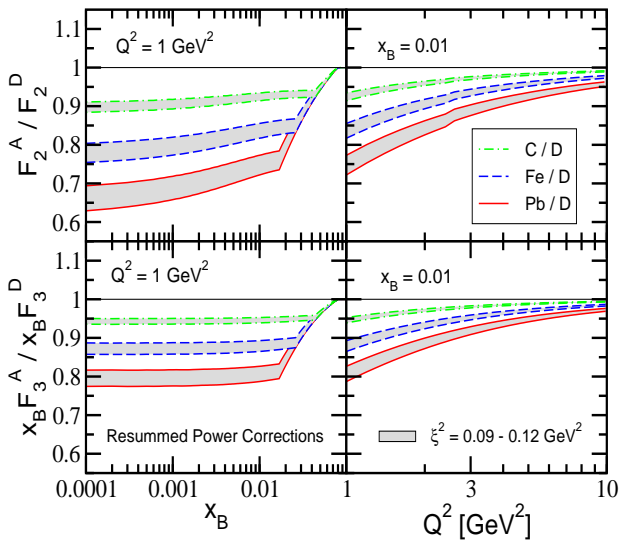}
\vspace*{-.in}
\caption{ The predicted nuclear modification for isoscalar-corrected
$^{12}C, \, ^{56}Fe$ and  $^{208}Pb$ to the neutrino-nucleus 
DIS stricture functions $F^A_2(x_B,Q^2)$ (top) and $x_B F^A_3(x_B,Q^2)$ 
(bottom) versus Bjorken $x_B$ (left) and $Q^2$ (right). The bands 
correspond to $\xi^2 = 0.09-0.12$~GeV$^2$~\cite{Qiu:2003vd}. }
\label{fig-ratios}
\end{center} 
\vspace*{-4mm}
\end{figure}

Figure~\ref{fig-ratios} shows the modification to the 
DIS structure functions from Eqs.~(\ref{F13resWp})-(\ref{FLresWn}) 
for large nuclei ($^{12}C, \, ^{56}Fe$ and $^{208}Pb$)   relative to 
the deuteron, calculated with the CTEQ6L parton distribution 
functions~\cite{Pumplin:2002vw}.
The bands correspond to a scale for power corrections 
$\xi^2 = 0.09 - 0.12$~GeV$^2$, extracted from the analysis~\cite{Qiu:2003vd}
of the NMC and E665 data~\cite{Arneodo:1995cs}. The transition 
$0.01 \leq x_B \leq 0.1$ region represents the onset of 
coherent interactions, Eq.~(\ref{uncertain}), and the modest 
$x_B$-dependence for $x_B \leq 0.01 $  is driven by the change in the 
local slope of the PDFs. The right panels 
show the $Q^2$ dependence of the  modification 
from the resummed  nuclear-enhanced power corrections, which is 
noticeably stronger than the DGLAP evolution of leading 
twist shadowing in the nPDFs~\cite{Eskola:1998df}. 
The difference in the suppression pattern of $F_2^A$ and $x_B F_3^A$  
in Fig.~\ref{fig-ratios} is qualitatively described by 
Eq.~(\ref{analyt}). In contrast, it has been suggested in 
the framework of a Gluber-Gribov approach~\cite{Kulagin:1998sv} 
that the suppression of the non-singlet distribution may be 
significantly larger than of the singlet one
($R^{A/A^\prime}_{sea/val.} > 1$ for $A > A^\prime$). 
Such distinctly different predictions should be testable 
in the future $\nu$-Factory experiments, 
for example  at the Fermilab NuMI facility~\cite{Morfin:vh}.

Although the current $\nu(\bar{\nu})$-$A$ DIS measurements are
mostly on nuclear targets \cite{Yang:2001xc}, these data 
 lack the necessary 
atomic weight systematics to identify small-$x_B$ nuclear shadowing. 
We do, however,  note that our results provide a consistent explanation 
of the observed small- and moderate-$Q^2$ power law deviation at 
small-$x_B$  of the preliminary NuTeV data~\cite{Naples:2003ne} on 
$F_2^A(x_B,Q^2)$ and  $x_B F_3^A(x_B,Q^2)$ from  
the next-to-leading  order leading twist QCD predictions 
using MRST parton distribution functions \cite{Thorne:1997uu}. 

%
%

The latest global QCD fits include $\nu(\bar{\nu})$-$A$ DIS data 
without nuclear correction other than isospin \cite{Pumplin:2002vw}.  
Such analysis would tend to artificially  eliminate most of the higher 
twist contributions discussed here due to a trade off between the power 
corrections in a limited range of $Q^2$ and the shape of the fitted 
input distributions at $Q_0^2$, especially within the error bars
of current data.
An effective way to verify the importance of the nuclear enhanced 
power corrections for neutrino-nucleus deeply inelastic scattering  
is via the QCD  sum rules, in particular, the Gross-Llewellyn Smith (GLS) 
sum rule~\cite{Gross:1969jf}
\begin{equation}
S_{\rm GLS} 
= \int_0^1 dx_B \,\frac{1}{2x_B} 
\left( x_B F_3^{\nu A}+ x_B F_3^{\bar{\nu} A} \right)\; . 
\label{Sgls}
\end{equation}
At tree level  Eq.~(\ref{Sgls}) counts the number of valance quarks 
in a nucleon, $S_{\rm GLS} = 3$.  Since valence quark number 
conservation is enforced in the extraction of twist-2 nucleon/nucleus
PDFs, the adjustments of input parton distributions can alter 
their shape  but not the numerical contributions to the GLS sum rule.

The effect of scaling violations can modify  $S_{\rm GLS}$, 
and at ${\cal O}(\alpha_s)$ \cite{Brock:1993sz}
\begin{equation}
\Delta_{\rm GLS} 
\equiv  \frac{1}{3}\left(3 - S_{\rm GLS}\right)
= \frac{\alpha_s(Q^2)}{\pi} + \frac{{\cal G}}{Q^2} 
+{\cal O}(Q^{-4}) \, .
\end{equation} 
Loop contributions to the GLS sum rule are known to 
${\cal O}(\alpha_s^3)$~\cite{Hinchliffe:1996hc}. 
Although power corrections can also modify the shape of {\em nucleon}
structure functions, recent  precision DIS data on both hydrogen and
deuterium targets from JLab \cite{JLab:data} indicate that effects from 
higher twist
to the lower moments of structure functions are very small at 
$Q^2$ as low as 0.5 GeV$^2$, which confirms 
the Bloom-Gilman duality \cite{Duality}.  
A recent phenomenological study~\cite{Alekhin:2002fv} 
also suggests that power corrections to the proton 
$F_2(x_B,Q^2)$ have different sign in the small- and large-$x_B$
regions and largely cancel in the QCD sum rules.



\begin{figure}[t!]
\begin{center} 
\hspace*{0in} 
\psfig{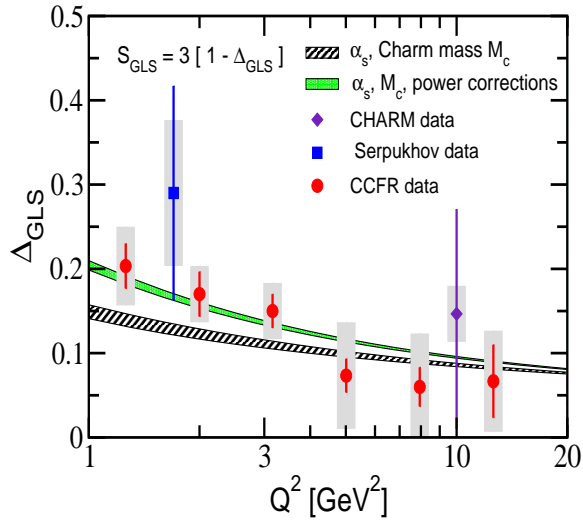}
\caption{Top panel: $\Delta_{\rm GLS}$ calculated to ${\cal O}(\alpha_s)$
with charm mass ($M_c=1.35$~GeV) effects (stripes) and $M_c$ +
resummed power corrections (band).
Data is from CCFR, CHARM and IHEP-JINR~\cite{Kim:1998ki}.
}
\label{sum-rules}
\end{center} 
\vspace*{-2mm}
\end{figure}

On the other hand, the coherence between different nucleons 
inside a large nucleus is only relevant 
for $x_B\leq x_N$. The suppression of 
structure functions at small Bjorken $x_B$ in Fig.~\ref{fig-ratios},
caused by the nuclear enhanced dynamical power corrections, 
cannot be canceled in the moments and  further reduces  the 
numerical value of $S_{\rm GLS}$. Figure~\ref{sum-rules} shows a 
calculation of $\Delta_{\rm GLS}$ from Eqs.~(\ref{F13resWp}) and  
(\ref{F13resWn}) for $^{56}Fe$.  
While the effect of charm mass is seen to be small relative to
$\alpha_s/\pi$, for $Q^2 \sim 1$~GeV$^2$  nuclear enhanced higher
twists may contribute as much as $\sim 10\%$ to 
$\Delta_{\rm GLS}$. Their $Q^2$ behavior is 
consistent with the trend in the current data~\cite{Kim:1998ki}.
For comparison, we found no deviations for this sum rule
induced by the EKS98 scale dependent parameterization of 
nuclear effects~\cite{Eskola:1998df}, which is again a consequence of 
the valance quark number conservation in leading twist shadowing.

\section{Implications for extraction of $\sin^2 \theta_W$}

Based on a comparison of charged and neutral current neutrino 
interactions  (separated on the basis of event topology) 
with an iron-rich heavy target, the NuTeV collaboration
reported a measurement of 
$$\sin^2 \theta_W^{({\rm on-shell })} = 0.2277 \pm 0.0013({\rm stat.}) 
 \pm 0.0009({\rm syst.})  \;,$$ 
neglecting the very small top quark and Higgs mass corrections.  
This result is approximately 3 standard deviations \cite{Zeller:2001hh} 
above the  Standard Model (SM) expectation value  
$\sin^2 \theta_W = 0.2227 \pm 0.0004$.
The NuTeV's result was derived from a quantity that is a close 
approximation to the Paschos-Wolfenstein relationship
\begin{equation} 
R^-=\frac{\sigma^{\rm NC}(\nu)-\sigma^{\rm NC}(\bar{\nu})}
         {\sigma^{\rm CC}(\nu)-\sigma^{\rm CC}(\bar{\nu})}
   \approx \frac{1}{2} -\sin^2\theta_W\, .
\label{pw-ratio}
\end{equation}
Corrections to  Eq.~(\ref{pw-ratio}) include higher order 
and nonperturbative QCD effects, higher order electroweak effects and
nuclear effects. 
Based on a QCD global analysis of parton structure, 
Kretzer et al. argue \cite{Kretzer:2003wy} that the
uncertainties in the theory which relates $R^-$  to  $\sin^2\theta_W$ are
substantial on the scale of the precision 
NuTeV data and suggest that the $\sin^2\theta_W$ measurement, NuTeV dimuon
data and other global data sets used in QCD parton structure analysis
can all be consistent within the SM.

Because a  heavy target was used, several nuclear effects can enter
the cross sections to influence the extraction of $\sin^2 \theta_W$
\cite{McFarland:sk}. 
Since nuclear enhanced power corrections were not included in NuTeV's
analysis, Miller and Thomas pointed out that nuclear 
shadowing from a vector meson dominance (VMD) model could 
affect the charged and neutral current neutrino scattering
differently, and therefore change the predictions for the ratios of
neutral current (NC) over charged current (CC) cross sections,
$R^{\nu(\bar{\nu})} 
=\sigma^{\rm NC}(\nu(\bar{\nu}))/\sigma^{\rm CC}(\nu(\bar{\nu}))$, 
and the extraction of $\sin^2 \theta_W$ \cite{Miller:2002xh}.  
The NuTeV collaboration argued \cite{Zeller:2002et,McFarland:sk} 
that such possibility was considered unlikely because 
$R^-$ has little sensitivity 
to process-dependent nuclear effects.

In this letter we calculated the process-dependent nuclear effects
in the neutrino-nucleus {\it differential} cross sections,
Eq.~(\ref{diffdis}), in the {\em perturbatively accessible} 
DIS region.  Our predictions on the nuclear modification to the 
$\nu (\bar{\nu})$-$A$  structure functions in Fig.~\ref{fig-ratios} 
should be relevant for $Q^2$ between 1 and 10~GeV$^2$.  
While for the mean  $\langle Q^2 \rangle_\nu = 25.6$~GeV$^2$
and $\langle Q^2 \rangle_{\bar{\nu}} = 15.4$~GeV$^2$ 
the effect of dynamical power corrections 
is small, a large fraction of the final data sample  cover the 
$x_B< 0.1, Q^2 < 10\; {\rm GeV}^2 $ range where shadowing can be as 
large as $ \sim 20 \% $. 
We note that the NuTeV measurement constitutes a  $\sim 2\%$  
{\it increase} in the value of $\sin^2 \theta_W$ relative to the SM, 
or equivalently $\sim 4\%$ {\it reduction} of the expected total
neutrino-nucleus cross section.  Including  shadowing into the
expected total cross sections will certainly reduce the discrepancy
of  $\sin^2\theta_W$.  However, without knowing the nuclear
enhanced power corrections to the structure functions at $Q^2<1$~GeV$^2$, and 
the detailed Monte Carlo simulation of event distributions, it is
difficult to estimate the precise corrections to the
extraction of $\sin^2\theta_W$.  
We, nevertheless, note that at small Bjorken $x_B$, the calculated
nuclear structure functions  $F_2^{A}(x_B,Q^2)$ and $F_3^{A}(x_B,Q^2)$ 
in neutrino-iron DIS qualitatively describe the low-$x_B$ 
and low-$Q^2$ suppression trend in the preliminary data, 
presented by the NuTeV collaboration at DIS 2003 
\cite{Naples:2003ne}.

\section{Conclusions}

In the framework of the perturbative QCD collinear factorization approach 
\cite{Collins:gx,Qiu:xy}, we computed and resumed the tree level 
perturbative expansion of nuclear enhanced power corrections to
the structure functions measured in inclusive (anti)neutrino-nucleus 
deeply inelastic scattering. 
We demonstrated that these corrections commute with the final state 
heavy quark effects and identified the new contributions to the 
longitudinal  structure function $F_L^A(x_B,Q^2)$. 
Our calculated $Q^2$-dependent modification to the Gross-Llewellyn Smith
sum rule 
agrees well with the existing measurements on an iron 
target~\cite{Kim:1998ki}. Our  
approach predicts a 
non-negligible difference in the small-$x_B$ shadowing 
of the structure functions $F_2^A(x_B,Q^2)$ ($F_1^A(x_B,Q^2)$) 
and $F_3^A(x_B,Q^2)$, which is consistent with the trend in the 
preliminary NuTeV data
\cite{Naples:2003ne}.  Although our results, valid in the
perturbative region, are unlikely to have an immediate impact on 
the NuTeV's extraction of $\sin^2 \theta_W$, the predicted $x_B$-,
$Q^2$-, and $A$-dependence of the structure functions in the
shadowing region can be tested  at the future Fermilab 
NuMI facility \cite{Morfin:vh}.

\begin{acknowledgments}
This work is supported in part by the US Department of Energy  
under Grant No. DE-FG02-87ER40371. We thank G. Zeller, 
J. Morfin, G. Sterman and E. Shuryak for useful discussion.
\end{acknowledgments}


\end{document}